# Superfluid currents in half-moon polariton condensates

E. S. Sedov[1,2,3*], V. A. Lukoshkin[4,5], V. K. Kalevich[4,5], Z. Hatzopoulos[6], P. G. Savvidis[1,2], A. V. Kavokin[1,2,5]

[1]Westlake University, 18 Shilongshan Road, Hangzhou 310024, Zhejiang Province, China
[2]Institute of Natural Sciences, Westlake Institute for Advanced Study, 18 Shilongshan Road, Hangzhou 310024, Zhejiang Province, China
[3]Vladimir State University named after A. G. and N. G. Stoletovs, 87 Gorky str., Vladimir 600000, Russia
[4]Ioffe institute, Russian Academy of Sciences, 26 Politekhnicheskaya, St-Petersburg 194021, Russia
[5]Spin Optics laboratory, St-Petersburg State University, 1 Unlianovskaya, St-Petersburg 198504, Russia
[6]FORTH, Institute of Electronic Structure and Laser, 71110 Heraklion, Crete, Greece

*evgeny_sedov@mail.ru

**Abstract**

We excite exciton-polariton condensates in half-moon shapes by the non-resonant optical excitation of GaAs-based cylindrical pillar microcavities. In this geometry, the $\pi$-jump of the phase of the condensate wave function coexists with a gradual $\pm\pi$ phase variation between two horns of the half-moon. We switch between clockwise and counter-clockwise phase currents by slightly shifting the excitation spot on the surface of the pillar. Half-moon condensates are expected to reveal features of two-level quantum systems similar to superconducting flux qubits.

# Introduction

The orbital angular momentum (OAM), that is an extrinsic degree of freedom of a stationary light beam is in the focus of our study. In the last couple of decades, OAM of light attracts much interest as a prospective entity for application in quantum computation and communication [1, 2]. The average OAM being a collective degree of freedom, i.e. characterizing a macroscopic (multi-particle) quantum state, offers a remarkable advantage over single-particle characteristics for information storage and processing. It is topologically protected against decoherence processes including those caused by dissipation, scattering and noise effects [3]. The loss or dephasing of individual photons does not lead to the loss of spatial coherence of the whole state.

Among the most well studied states with non-zero OAM are vortex states characterized by the azimuthal phase dependence of the form $\exp(im\theta)$, where the integer $m$ is a quantum number frequently referred to as a topological charge [4], and $\hbar m$ is the OAM per photon [5], $\theta$ is the azimuthal angle. There exist several approaches for generation of states with non-integer orbital momenta based on the introduction of local phase jumps that make the phase variation in the remaining part of the closed loop different from a multiple of $2\pi$. Depending on the physical system under consideration, various approaches are used to create quantum states with the non-integer OAM. For generation of optical fractional vortices one needs specially designed optical elements, fractional vortex lenses or spiral phase plates [6,7]. In superfluid ring-shaped Bose-Einstein condensates of ultracold atoms, the method in demand for generating flow states characterized by a phase slip consists in the insertion of a so-called weak link [8,9]. A weak link represents a depletion in the condensate density produced by the external potential barrier moving along the circumference of the condensate at a given velocity. It induces back-scattering for the condensate flows spreading clockwise and counter-clockwise along the ring. This results in interference patterns of the scattered flow states. In recent years, a significant progress in shaping of Bose-Einstein condensates of light-matter quasiparticles, exciton-polaritons, has been reported [10, 11]. In contrast to atomic condensates, polariton condensates may be generated at relatively high temperatures, including the room temperature [12] that opens the way to their application in classical and quantum information processing.

In the present work, creation of polariton phase currents characterised by fractional topological charges is demonstrated by the near-field optical interferometry in cylindrical micropillar microcavities. We experimentally study semiconductor micropillars where the demonstration of concentric circular and lobe condensate patterns [13] and generation of persistent circular currents [14] were reported earlier. We introduce a weak link in a circular polariton condensate that takes a half-moon shape. The polariton density depletion between the horns of the half-moon is provoked by the repulsive interaction of the condensate polaritons with the reservoir of incoherent excitons generated by a non-resonant optical excitation under the optical pump spot. We observe clock-wise and anti-clockwise persistent circular current states characterised by half-integer orbital momenta. This observation shows the high potentiality of half-moon polariton condensates for applications in the quantum information processing as is explained below.

We refer here to the theoretical work [15] that uses a quantum mechanical approach for the description of fractional OAM states. In that model, the fractional OAM state is considered as a *qunit*, a quantum state in $N$ (in theory, infinite) dimensions [16], representing a superposition of current states with integer winding numbers $m$ (OAM per state). A convenient approach consists in the reduction of the infinite Hilbert space of the system to the two-dimensional subspace, where the fractional OAM state can be considered as a *qubit* in the $(+m, -m)$ basis [17]. Half-moon polariton condensates offer an opportunity to single out the pair of vortex states characterized by OAM of $+1/2$ and $-1/2$. This pair of states may be used as a basis for a polariton qubit that is similar to the well-known flux qubits based on superconducting circuits with embedded Josephson junctions [18].

## Polariton condensate formation.

Schematically the creation of the half-moon polariton condensate is illustrated in Fig. 1(a). Figure 1(b) presents the energy transfer scheme in the system. The sample under examination is a cylindrical pillar of a diameter of 25 $\mu$m etched from a planar 5$\lambda$/2 AlGaAs distributed Bragg reflector microcavity with an ensemble of embedded quantum wells. The measured cavity quality factor is $Q \approx 16000$. Due to the strong coupling of quantum well excitons with cavity photons confined in the structure growth direction, the structure supports two new eigenmodes split in energy. These are exciton polariton modes that form the upper and lower dispersion branches (pink and light blue curves in Fig. 1(b), respectively). For creating a condensate of the lower branch polaritons, we use the non-resonant pumping scheme [14]. The linearly polarized optical cw pump beam coming from the Ti:sapphire laser and being characterized by an energy far above the exciton resonance (about 110 meV above the minimum of the lower polariton branch) creates a reservoir of incoherent excitons. Due to the stimulated scattering, the reservoir excitons relax to the polariton state at the bottom of the lower dispersion brunch. Since at the non-resonant pumping, the spin and orbital characteristics of light do not transfer to the ground polariton state directly, we do not imprint any specific phase to the condensate with the pump laser.

The condensate of polaritons in a pillar is described by a single many-body wave function $\Psi(\mathbf{r})$. If the pump beam is coaxial with the pillar, the rotationally symmetric ring-shaped ground polariton state is formed with a uniform phase that does not vary in the area of the pillar, see Ref. [13]. The exciton cloud localized under the pump beam creates an effective potential for polaritons repelling them from the centre of the pillar. In a non-conservative system, the localized pump and spatially distributed losses cause polariton flows, which are characterized by the vector field $\mathbf{J} = (i/2)\left(\Psi\nabla\Psi^* - \Psi^*\nabla\Psi\right)$. The spontaneous or induced breaking of the rotational symmetry, e.g. due to the specific combination of the landscape of the static potential and the position of the pump spot, can lead to the formation of persistent ring current states of polaritons, characterized by the non-zero quantized OAM per particle [14]. The latter can be formally found from the vector flow field as $\ell = L_z/N$, where $L_z = \int(x\partial_y J_y - y\partial_x J_x)d\mathbf{r}$ is the actual OAM and $N = \int|\Psi|^2 d\mathbf{r}$ is the population of the condensate.

In the recent paper Ref. [19], some of the present authors proposed the concept of a polariton flux qubit. We have predicted theoretically that the introduction of a weak link in the circular polariton current pins a node of the condensate wave function leading to a jump of the condensate phase by $\pi$. The resulting current states are characterized by a fractional

OAM. In this Letter, we report on the experimental realization of the polariton current states with a fractional OAM in half-moon polariton condensates. The key difference in the geometry of our present experiment and the experiments reported in Refs. [13, 14] is the position of the pump spot. In contrast to the previous works, here we deliberately shift the pump beam from the centre of the pillar in such a way that the ring trap transforms to the C-shape one. The pump creates the potential barrier that plays the role of a weak link in a ring-shape superfluid circuit of exciton polaritons formed in the regime of polariton lasing. Figure 1(c) shows schematically the coupling of the counterpropagating currents via backscattering from the pump induced defect.

Although in a geometry characterized by the broken rotational symmetry OAM is not obliged to be integer, the overall variation of the condensate phase around the pillar still needs to be a multiple of $2\pi$. The introduced barrier acts as a phase delay line which imprints the phase jump within the condensate density gap. In Fig. 1(d), we schematically show the phase variation of the polariton current states, which correspond to the overall phase variation of zero, which forms the pair of lowest energy eigenstates of our system.

## Half-moon polariton condensates.

In Fig. 2, we present the experimental evidence for the formation of a polariton condensate characterized by a fractional OAM $\ell > 0$. The pronounced crescent shape of the steady state polariton density in Fig. 2(a) is a result of the displacement of the pump beam by a sub-micrometer distance to SW from the center of the pillar. In the Supporting Information, we show that despite of the apparent complicated shape of the polariton density, the latter can be safely separated into the azimuthal and radial components.

To reveal the phase structure of the half-moon polariton condensate, we have studied the interference pattern formed by the photoluminescence signal emitted by the condensate with the coherent spherical reference wave of the same frequency, see Fig. 2(b). The reference wave was obtained by magnifying the peripheral part of the condensate emission and guiding it through a converging lens spaced from the image plane by the distance exceeding its focal length. The smooth variation of the phase of the polariton condensate in the most part of the perimeter of the pillar results in the appearance of concentric fringes in the interferogram. The smooth patterns are broken in the vicinity of the deep in the polariton density. The observed abrupt shift of the interference fringes is caused by the jump of the phase between the horns of the half-moon condensate. Figures 2(c)–2(e) confirm this observation. Adapting the extended Fourier-transform method described in Ref. [14] for the case of closed-fringe patterns, we extract from the interferogram in Fig. 2(b) the phase of the condensate emission relative to that of the reference wave, see Fig. 2(c). The variation of the phase of the polariton condensate along the closed pass on the crest of the condensate density is shown in Fig. 2(d). The full reconstruction of the phase of the half-moon condensate $\varphi(r, \theta)$ is shown in Fig. 2(e). One can see that the anticipated phase jump by approximately $+\pi$ between the horns of the half-moon is unambiguously observed. Herewith, the phase decreases roughly linearly with the increase in the azimuthal angle outside the condensate density deep.

To visualise the polariton phase current in the half-moon condensate, in Fig. 2(f), we plot the vector field **J**. To extract it from near-field photoluminescence and interferometry images, we represent the polariton condensate wave function as $\Psi(r, \theta) = n^{1/2}(r, \theta) \exp[i\varphi(\theta)]$, see the Supporting Information for details of the fitting procedure for the polariton density $n(r, \theta)$ and the phase $\varphi(\theta)$. The reconstructed wave function allowed us to estimate the OAM per particle for the observed current state as $\ell \approx +0.5$. The vector field winds around the centre of the pillar. However, in contrast to the current states with an integer OAM (e.g. in Ref. [14]), the magnitude of the field of the fractional OAM state varies as a function of the azimuthal angle. It drops down to zero with the decrease of the polariton density.

Experimentally, we were also able to obtain the half-moon condensate with a negative OAM, $\ell < 0$, see Fig. 3. To do this, we shifted the pump beam to NW from the centre of the pillar. In this case, the observed polariton density variation between the horns of the condensate is sharper than in Fig. 2(a). Nevertheless, the azimuthal density distribution is well described by the same analytical function with different fitting coefficients. As for the radial component of the condensate wavefunction, the discrepancy in the shapes of the functions that provide the best fit for the two experiments is less than one percent, so that it can safely be neglected, see the Supporting Information. In a full similarity to the case of $\ell \approx +0.5$ condensate analyzed above, we performed the interferometry measurements followed by the restoration of the phase of the condensate, see Figs. 3(b)–3(e). The jump of the phase by approximately $-\pi$ is apparent on the angular dependence of the condensate phase $\varphi(\theta)$. Outside the break point, the dependence $\varphi(\theta)$ is non-linear and it differs from one shown in Figs 2(d)–2(e). Still, its monotonic increase with the increase of $\theta$ away from the condensate density deep is an unambiguous proof of a counter-clockwise phase current.

In Fig. 3(f) we plot the polariton flow field $\mathbf{J}$ using the reconstructed wave function of the $\ell < 0$ state. Due to the more pronounced inhomogeneity of the considered OAM state, the density of field lines is significantly higher near the maximum of the polariton density than near the polariton density deep. Nevertheless, despite of such a difference in the flow field structure, the two states are characterized by almost similar absolute values of the OAM per particle: for the $\ell < 0$ condensate state, we estimate the OAM per particle as $\ell \approx -0.5$.

Both the density and the phase structures of the observed half-moon polariton condensates are very stable. Once emerged in the cw experiment, the condensate density distribution or the broken spiral interference pattern remain unchanged during the observation.

## Discussion.

Half-quantum circulation states of polariton condensates have been known since the end of the last decade. [20–24]. Chronologically, the first such state known as a half-vortex state has been proposed in Ref. [20] and then observed in Refs. [21–23]. It appears in a spinor polariton liquid in a planar microcavity and is characterized by a vortex state in one of circular polarizations and no vortex in another circular polarization. For a fixed radius with respect to the half-vortex core, a linear polarization plane rotates with the change of the azimuthal angle $\theta$ around the core. For $\theta$ varying by $2\pi$, the linear polarization plane rotates by $\pi$. Herewith, the contribution of the circular polarization is independent of the azimuthal angle. Another half-quantum circulation state has been experimentally demonstrated in Ref. [24] recently. Existing only in a ring geometry, this state also provides a $\pi$ rotation of the linear polarization plane, which, however, is necessarily accompanied by a flip of the circular polarization upon a turnover along the ring. The common feature of the discussed half-quantum circulation states is that the polarization degree of freedom is involved.

In order to distinguish the observed half-quantum orbital states of half-moon condensates from the discussed states [20–24], we have analyzed the polarization-resolved near-field photoluminescence spectra of our system. The experimentally measured spatially resolved polarization (Stokes vector) components, $S_x$, $S_y$ and $S_z$, are presented in Fig. 4 for the states $\ell \approx 0.5$ [panels (a)–(c)] and $\ell \approx -0.5$ [panels (e)–(g)]. To extract the Stokes vector components, $S_x = (I_\leftrightarrow - I_\updownarrow)/(I_\leftrightarrow + I_\updownarrow)$, $S_y = (I_\nwarrow - I_\nearrow)/(I_\nwarrow + I_\nearrow)$ and $S_z = (I_\circlearrowright - I_\circlearrowleft)/(I_\circlearrowright + I_\circlearrowleft)$, we have measured intensities of six components of the condensate photoluminescence: linearly polarized horizontally ($I_\leftrightarrow$) and vertically ($I_\updownarrow$), along diagonal ($I_\nwarrow$) and perpendicular to diagonal ($I_\nearrow$), and right-hand ($I_\circlearrowright$) and left-hand ($I_\circlearrowleft$) circularly polarized. In Figs. 4(d) and 4(h), we show the polarization ellipses characterizing the polarization of polaritons in different positions along the ridge of the polariton condensate. These polarization measurements allow us to rule out half-quantum circulation states as a possible origin of the observed fractional OAM of the polariton condensate, as we observe neither the rotation of the linear polarization plane no the flip of the circular polarization around the condensate circumference. It is clearly seen that the polariton density is almost homogeneous in all polarizations within the area of the pillar occupied by the condensate. The linear component $S_x < 0$ is the most pronounced in both experiments, herewith the angle or the linear polarization just slightly fluctuates when going around the pillar. The contribution of the circular polarization is weak, and the pattern of its fluctuations around the pillar can hardly be traced.

Both in our experiment and in the experiment discussed in Ref. [24], the polariton condensates in a ring geometry were studied. In Ref. [24], a harmonic potential formed by applying an inhomogeneous stress to the planar microcavity is in the basis of an annular trap. In our experiment, we used a micropillar to confine polaritons. The reservoir of nonresonantly pumped excitons was used in both experiments providing the central maximum of the confinement potential. The key difference in the geometries of the two experiments is that in Ref. [24] the annular potential is supplemented by the linear gradient in one direction so that polaritons exhibit an effect of an external force analogous to the gravitational force acting on massive particles. [25–27] It induces an inhomogeneity in the polariton condensate density and induces the symmetry breaking that manifests itself in polarization patterns. Although in our experiment the polarization degree of freedom does not affects the behavior of polaritons, nevertheless the rotational symmetry must be broken so that the circular polariton currents appear. In the azimuthally symmetric trap where the pump spot is placed in the center of the pillar, the symmetry can be either broken spontaneously leading to the appearance of ring circular polariton currents, or preserved that is manifested in the formation of the current-less ring-shaped polariton state. The manipulation by OAM in the experiments in Figs. 2 and 3 was performed by shifting of the pump spot with respect to the center of the pillar. However the rotational symmetry is not broken by the shift itself. The chirality in the system is introduced by the combination with the asymmetry of the stationary potential, e. g. due to the slight ellipticity of the pillar or due to possible defects or inhomogeneities. In the recent paper [14], we successfully used this method to generate circular current states of polaritons with integer OAM. In the present manuscript, shifting of the pump spot is aimed not only to breaking the azimuthal symmetry but also to switching from the ring geometry to the half-moon geometry. In the latter case, the exciton reservoir acts as a potential barrier, which induces a deep in the density [28] and perturbs the phase of the polariton condensate.

The considered polariton system forms a robust optically controlled solid-state platform for implementation of a quantum register. The discussed half-integer OAM states representing a macroscopic superposition of current states are valuable candidates for the realisation of flux qubits that would play a role of the building blocks of the register.

If the rotational symmetry of potential is preserved, OAM of the polariton condensate in a pillar holds integer values [14]. Herewith, the different OAM states (including the stationary state with $\ell = 0$) are separated in energy from each other by the Galilean transformation due to the variation of the phase around the pillar centre [29]. The inclusion of the weak link, regardless of its origin, breaks the Galilean invariance and leads to the appearance of a node in the condensate wave-function. The newly established polariton flow eigenstates are characterized by half-integer OAM, that makes them analogous to the clockwise and counter-clockwise current states formed by a swarm of Cooper pairs in a superconducting circuit with an embedded Josephson junction [30–32]. Circuits of superfluid polariton currents demonstrate a very similar physics to superconducting circuits used in flux qubits. Half-integer orbital momenta states of a polariton condensate demonstrated here may be considered as two projections of the polariton flux qubit. Superposition states of clockwise and counter-clockwise phase currents would be characterized by oscillating trajectories on a surface of the Bloch sphere as discussed in Ref. [19]. The unambiguous observation of robust half-quantum polariton currents of both signs reported here is a decisive step forward towards the realisation of a polariton flux qubit.

We would like to emphasize the advantages of the polariton flux qubits based on polariton half-moon condensates, over many other implementations of qubits. The orbital angular momentum of a condensate is a macroscopic degree of freedom. It is based on the collective coherent phenomenology of polariton currents flying along closed trajectories. In contrast to qubits based on intrinsic quantum states of a single particle, the OAM qubits are topologically protected against dephasing caused by losses and dephasing of individual quantum particles.

Polariton qubits also offer an advantage of a relatively simple optical read-out. This favourably distinguishes them from superconducting flux qubits, which require extremely sensitive magnetometers to minimize the perturbation to the qubit state when performing so called "weak measurements" for the read-out.

Moreover, the operational temperature of polariton flux qubits is expected to significantly exceed those of the best superconducting qubits. The experimental observation of stable room-temperature condensates of exciton polaritons [33, 34] paves the way to realisation of room-temperature polariton flux qubits.

# Acknowledgment

This work is from the Innovative Team of International Center for Polaritonics and is supported by Westlake University (Project No. 041020100118). E.S.S. acknowledges a partial support from the Grant of the President of the Russian Federation for state support of young Russian scientists No. MK-2839.2019.2. V.A.L., V.K.K. and A.V.K. acknowledge financial support from the Russian Foundation for Basic Research (RFBR Project No. 19-52-12032) and Saint-Petersburg State University for the research grant ID 40847559.

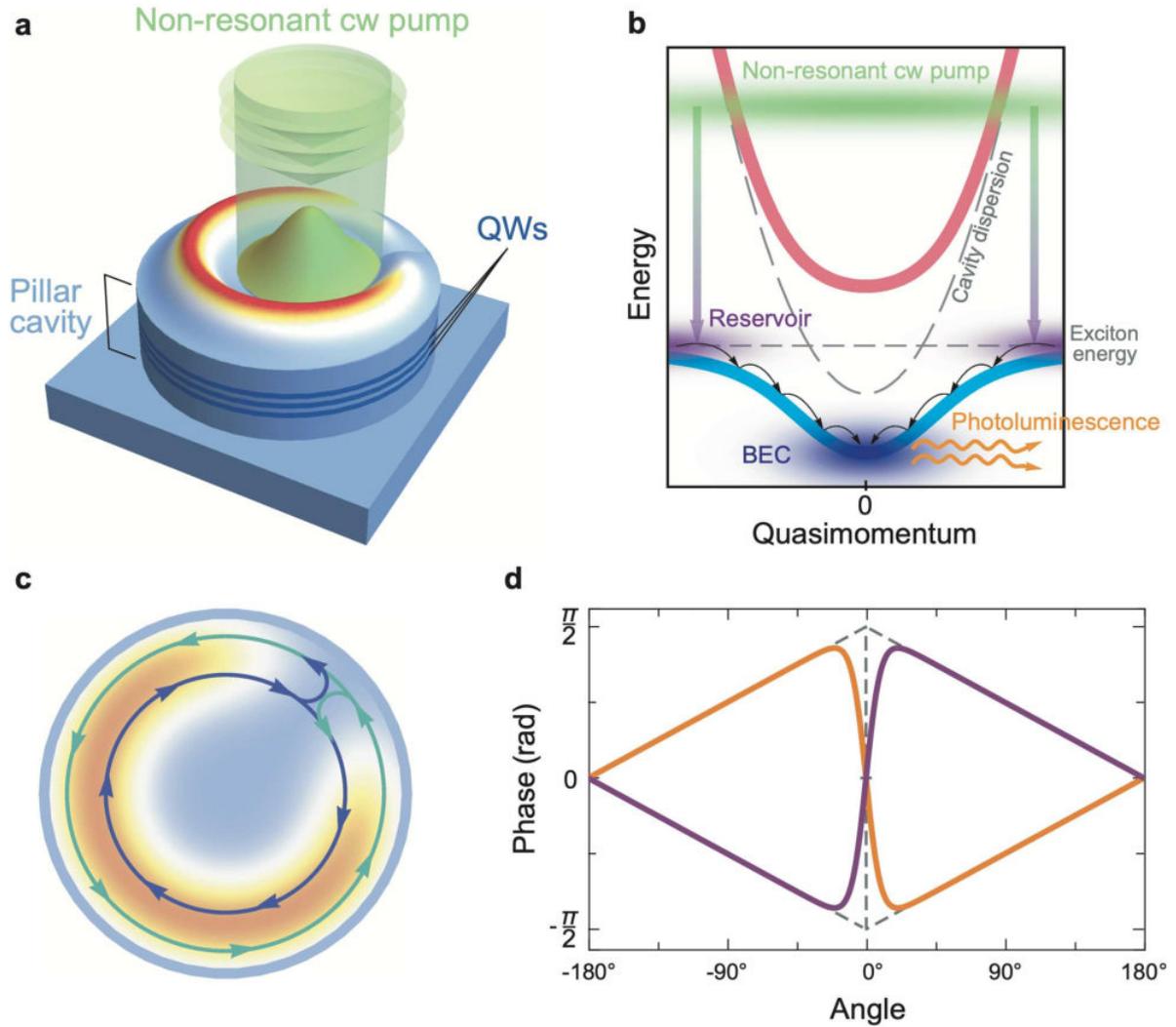

Figure 1: **Creation of a half-moon polariton condensate. a**, Schematic of the optical excitation of a half-moon polariton condensate. Polaritons are excited in a cylindrical micropillar cavity by a non-resonant cw Gaussian pump beam (pale green cylinder) slightly shifted from the centre of the micropillar. **b**, Schematic dispersion of exciton polaritons. The upper (pink curve) and lower (light blue curve) polariton dispersion branches result from the avoided crossing of the dispersions of excitons and photons (gray dashed curves) due to their strong coupling. The condensate of exciton polaritons (dark blue cloud) is fed from the reservoir of incoherent excitons (purple cloud) due to the exciton relaxation in energy via their interaction with phonons (black arrows). The reservoir is created by the optical cw pump (green cloud). Polaritons escape from the cavity as a photoluminescence (orange wavy arrows). **c**, Schematic of the coupling of polariton currents due to the backscattering from the polariton breach. **d**, The possible phase distributions in a half moon polariton condensate containing a node characterised by the phase jump of $-\pi$ (orange curves) and $+\pi$ (purple curves) at the weak link modelled by a smeared $\delta$-like barrier (the grey dashed curve corresponds to the limit of zero smearing).

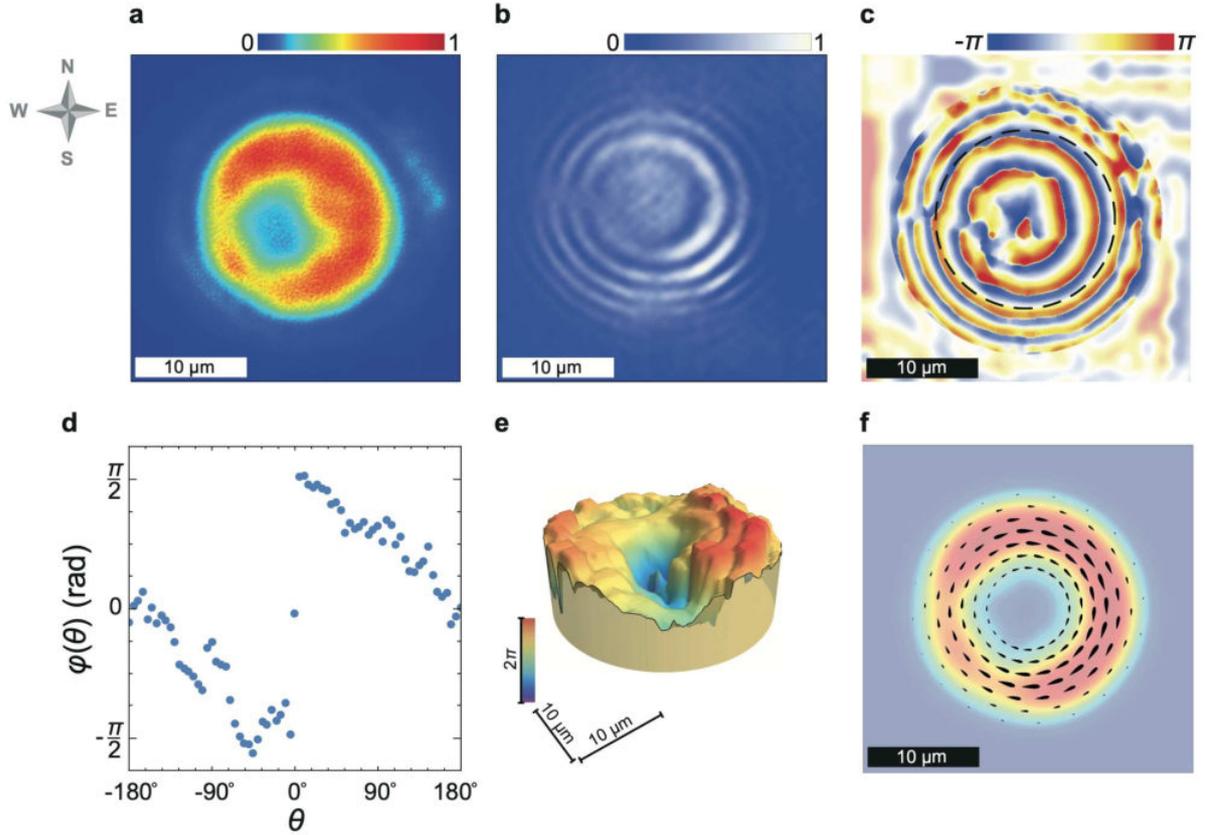

Figure 2: **The polariton current state with a positive fractional OAM. a**, The density distribution of the polariton condensate in the cavity plane. **b**, The interferometry image obtained from the interferometry of the photoluminescence of the condensate with the spherical reference wave. **c**, The relative phase map extracted from the interferometry image. **d**, Phase variation along the red dashed circumference in the panel **c**. The azimuthal coordinate of the phase jump is taken as $\theta = 0°$. **e**, Variation of the phase of the polariton condensate in the cavity plane. **f**, The vector polariton flow field **J** (black drops) extracted from the fitting of the experimental results by analytical functions, see Supporting Information for details. The background in the panel **f** is the fitted density distribution. The non-physical area outside the pillar image is shaded in the panel **c** and cut of in the panel **e**.

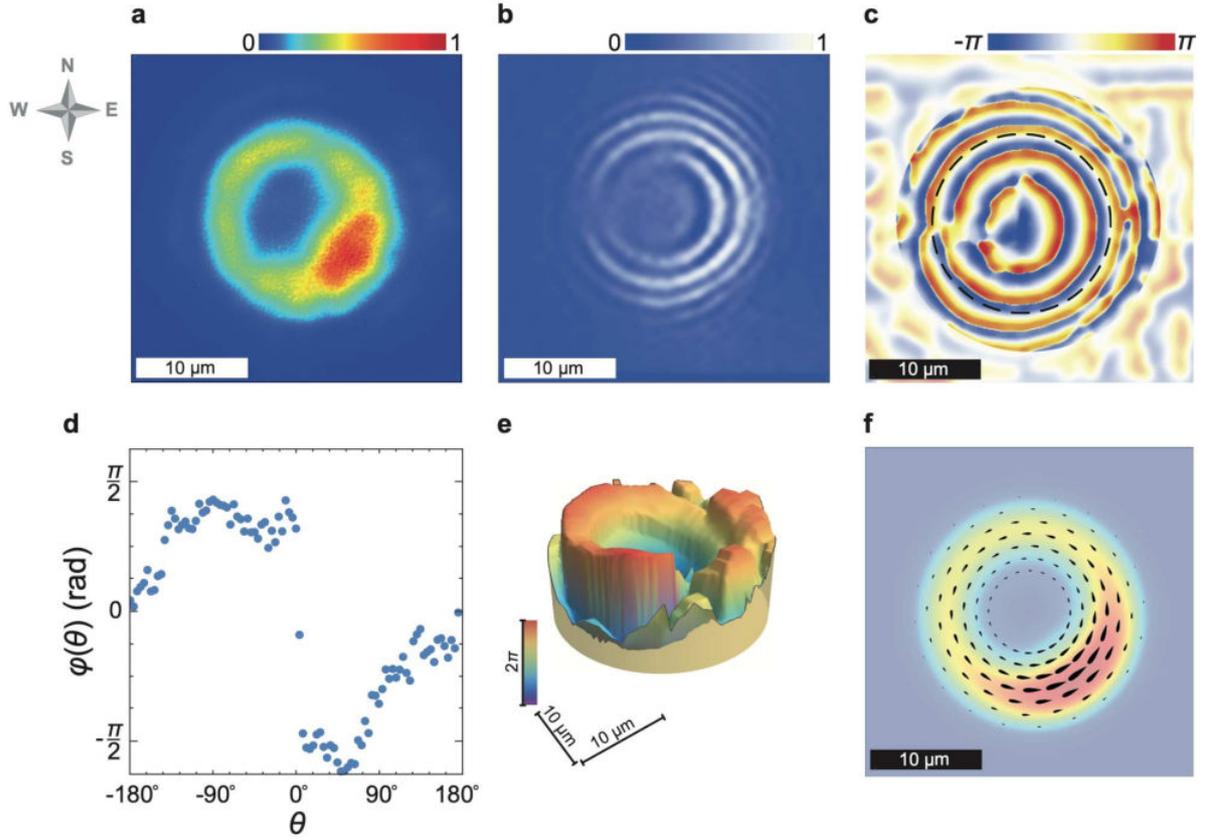

Figure 3: **The polariton current state with a negative fractional OAM. a**, The density distribution of the polariton condensate in the cavity plane. **b**, The interferometry image. **c**, The relative phase map extracted from the interferometry image. **d**, Phase variation along the red dashed circumference in the panel **c**. The azimuthal coordinate of the phase jump is taken as $\theta = 0°$. **e**, Variation of the phase of the polariton condensate in the cavity plane. **f**, The vector polariton flow field **J** (black drops) extracted from the fitting of the experimental results by analytical functions, see Supporting Information for details. The background in the panel **f** is the fitted intensity distribution.

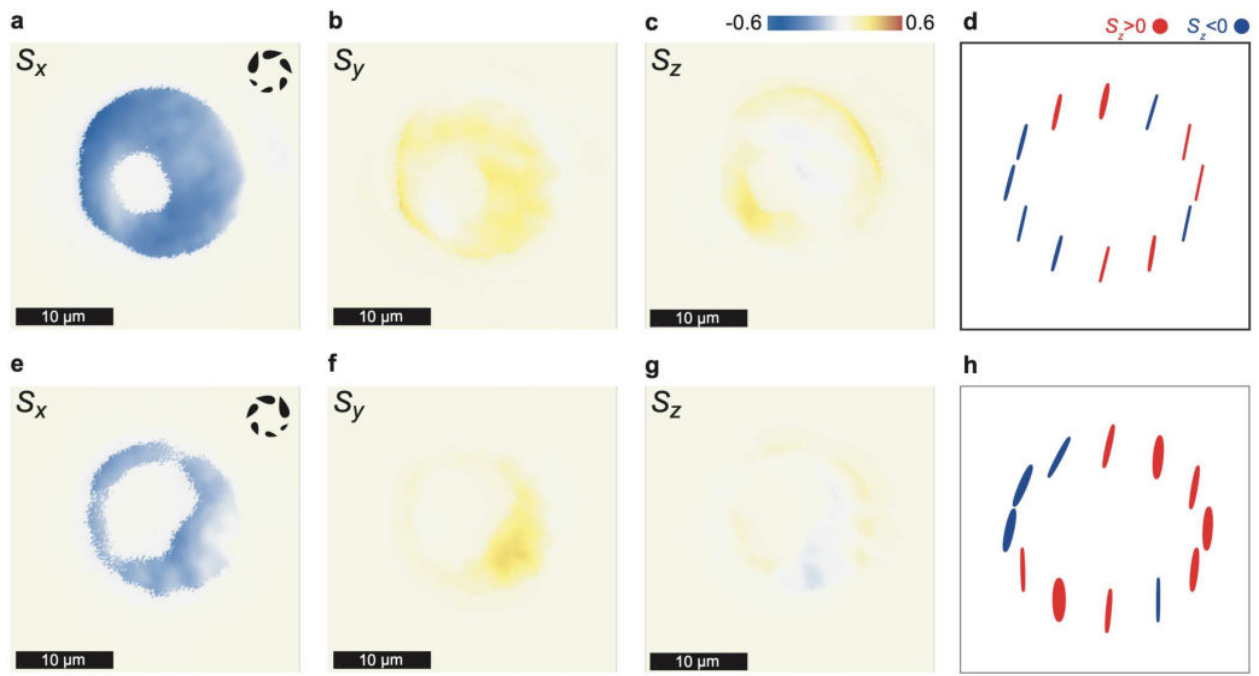

Figure 4: **Polarization of the half-moon condensates.** Maps of the **a**, **e**, linear, $S_x$, **b**, **f**, diagonal/antidiagonal, $S_y$, and **c**, **g**, circular, $S_z$, components of the polarization of light emitted by the condensates. **d**, **h**, Polarization ellipses characterizing the polarization of the emitted light at various positions along the condensate ring. Red and blue ellipses indicate $S_z > 0$ and $S_z < 0$, respectively. The upper and lower panels correspond to the counter-clockwise and clockwise polariton current states, respectively. The roundels of droplets in the upper right corners of panels **a** and **e** are given for clarity and indicate the polariton current state: counter-clockwise for $\ell > 0$ and clockwise for $\ell < 0$.

# Supporting Information: Superfluid currents in half-moon polariton condensates

E. S. Sedov, V. A. Lukoshkin, V. K. Kalevich, P. G. Savvidis, A. V. Kavokin

## S1 Details of the experimental setup and the sample

The experimental setup we used is described in detail in Ref. [1]. A cylindrical micropillar of a diameter of $25\,\mu$m is excited non-resonantly by a laser beam ($2\,\mu$m full width at half maximum) whose position is deliberately shifted from the center of the pillar. Both near-field photoluminescence and interferometry are studied in the polariton lasing regime. The interferometry images are obtained by the combining the image of the condensate with a magnified fragment of the same condensate that constituted a reference spherical wave in a Michelson interferometer. Polarization optical elements (linear polarizers and quarter- and half-wave plates) have been used to measure the spatially dependent components of the Stokes vector of light emitted by the condensate.

The sample is fully described in Ref. [1]. Four sets of three 10-nm GaAs quantum wells are placed at the antinodes of the cavity electric field to maximize the exciton-photon coupling. The studied $25\,\mu$m diameter pillar is characterized by a slight negative photon-exciton detuning of $\delta = -0.5$meV.

## S2 Fitting of the density and phase components of the half-moon condensates

First, we show that the azimuthal and radial components of the wave functions of the half-moon polariton condensates under consideration (in Figs. 2 and 3 in the main text) are separable and can be fitted independently. To this end, in the right panels of Fig. S1, we compare the radial distributions of the experimentally observed counter-clockwise (b) and clockwise (d) polariton flow states in different cross-sections, indicated in the left panels, (a) and (c), respectively. The figures show that for both condensates, distribution along the radius weakly depends on the azimuthal angle. The average normalized radial distribution $I(r) = C_r \int_0^{2\pi} n(r,\theta)d\theta$, where $C_r$ is the normalization coefficient, is very well fitted by the analytical function $F(r) = (e+dr+cr^2)e^{-a(r-b)^2}$. Values of the fitting coefficients are given in the caption to Fig. S1. Remarkably, the mismatch of the fitting functions for the counter-clockwise [$F_\frown(r)$] and clockwise [$F_\frown(r)$] condensates is less than one percent, $|(G_{\frown,\frown(\frown,\frown)} - G_{\frown,\frown})/G_{\frown,\frown(\frown,\frown)}| < 0.01$, where $G_{i,j} = \int_A F_i^{1/2}(r)F_j^{1/2}(r)rdr$ ($i,j = \frown,\frown$), $A$ is the numerical integration distance. The radial distribution of the polariton condensate is only determined by the shape and relative position of the external potential and effective potential due to the exciton repulsion which is similar for both experiments.

The normalized azimuthal intensity distribution $I(\theta) = C_\theta \int_A n(r,\theta)rdr$ of the counter-clockwise and clockwise half-moon polariton condensates discussed in the main text is shown in Figs. S2(a) and S2(b), respectively. Despite the apparent differences in the shapes of the distributions, both of them are very well fitted by the analytical function $F(\theta) = a\tanh[b(x+c)]^2 + a\tanh[d(x-e)]^2 + f$. Values of the fitting coefficients are given in the caption to Fig. S2.

Next, we fit the azimuthal variation of the phase of the half-moon polariton condensates, $\varphi(\theta)$. We use the reconstructed from the interferometry pictures phase variations in Figs. 2(d) and 3(d) in the main text as references. Figure S3 shows the fitting functions (black) of the form $a\theta + b\theta^2 + c\theta^3 + d\tanh(e\theta)$ in comparison with the reconstructed data for the positive [panel (a)] and negative [panel (b)] OAM per particle.

Finally, we can reconstruct the wave function of the half-moon condensate in the Madelung form $\Psi(r,\theta) = \sqrt{I(r)I(\theta)}$ $\times \exp[i\varphi(\theta)]$. Based on the reconstructed $\Psi(r,\theta)$, we can now estimate OAM per particle for the observed half-moon condensates as $\ell \approx 0.56$ for the experiment in Fig. 2 and $\ell \approx -0.54$ for the experiment in Fig. 3 of the main text.

## S3  Polarization properties of the half-moon condensate

To reveal the effect of the polarization of the pump beam on the polarization and orbital properties of polariton states, in a series of supplemental experiments, we have generated half-moon polariton condensates under the pump beams of three different polarizations: left-circular, right-circular and vertical. The upper row in Fig. S4 shows the interferometry images for the experiments demonstrating the dislocation of the interference fringes at the position of the phase jump. The panels in the lower row show polarization ellipses characterizing the polarization of the emitted light at various positions along the condensate ring for the discussed experiments.

As one can see, the condensate in the pillar does not inherit the pump beam polarization: regardless of the latter, the condensate polarization is linear with a slight ellipsity. The linear polarization angle barely deviates from the expectation value when going around the pillar. The percentage of the circular polarization is weak in all experiments. Remarkably, in these particular experiments, the ellipsity is even stronger in the case of the linearly polarized pump [Fig. S4(f)] rather than in the case of the circularly polarized pump [Fig. S4(d)]. The flip of the circular polarization does not follow any pattern. We believe the flip cannot significantly affect the topology and dynamics of superfluid polariton currents in the case of such a small ellipticity. The polarization behavior of the observed half-moon condensate clearly distinguishes it from that of other half-quantum circulation states [2–6].

## S4  The fractional OAM state as a superposition of current states

A current state with an arbitrary OAM in fact represents an infinite superposition of orthogonal states with integer winding numbers $m$ as follows [7]:

$$|\Psi\rangle = \sum_{m=-\infty}^{\infty} |\psi_m\rangle = \sum_{m=-\infty}^{\infty} a_m |m\rangle, \quad (1)$$

where $a_m$ is the weight coefficient of the current state $|m\rangle$; $\sum_{m=-\infty}^{\infty} |a_m|^2 = 1$. We demonstrated above separability of the condensate wave function in the radial [$\Psi_r(r)$] and azimuthal [$\Psi_\theta(\theta)$] components as $\Psi(r,\theta) = \Psi_r(r)\Psi_\theta(\theta)\exp[i\varphi(\theta)]$. Using the conservation of the radial components for the condensates, we decompose the polariton state $|\Psi\rangle$ in terms of the ring current states with the same radial distribution: $|m\rangle = \frac{1}{\sqrt{2\pi}}\Psi_r(r)\exp(im\theta)$. The weight coefficients $a_m$ are then found as follows [8]:

$$a_m = \frac{\int_0^{2\pi} \Psi_\theta(\theta)\exp\{i[\varphi(\theta) - m\theta]\}\,d\theta}{\left[2\pi\int_0^{2\pi} |\Psi_\theta(\theta)|^2 d\theta\right]^{1/2}}. \quad (2)$$

Figure S5 shows the contribution $|a_m|^2$ of the current states with the angular momenta $m$ to the discussed half-moon condensates. It is clearly seen that components with both positive and negative winding numbers contribute to the OAM per particle for both polariton states. The non-rotating state ($m = 0$) corresponding to the ground state of the system is present there. One should also note that the presented in the figure components from $m = -6$ to 6 cover more than 97% of the final state with the given OAM.

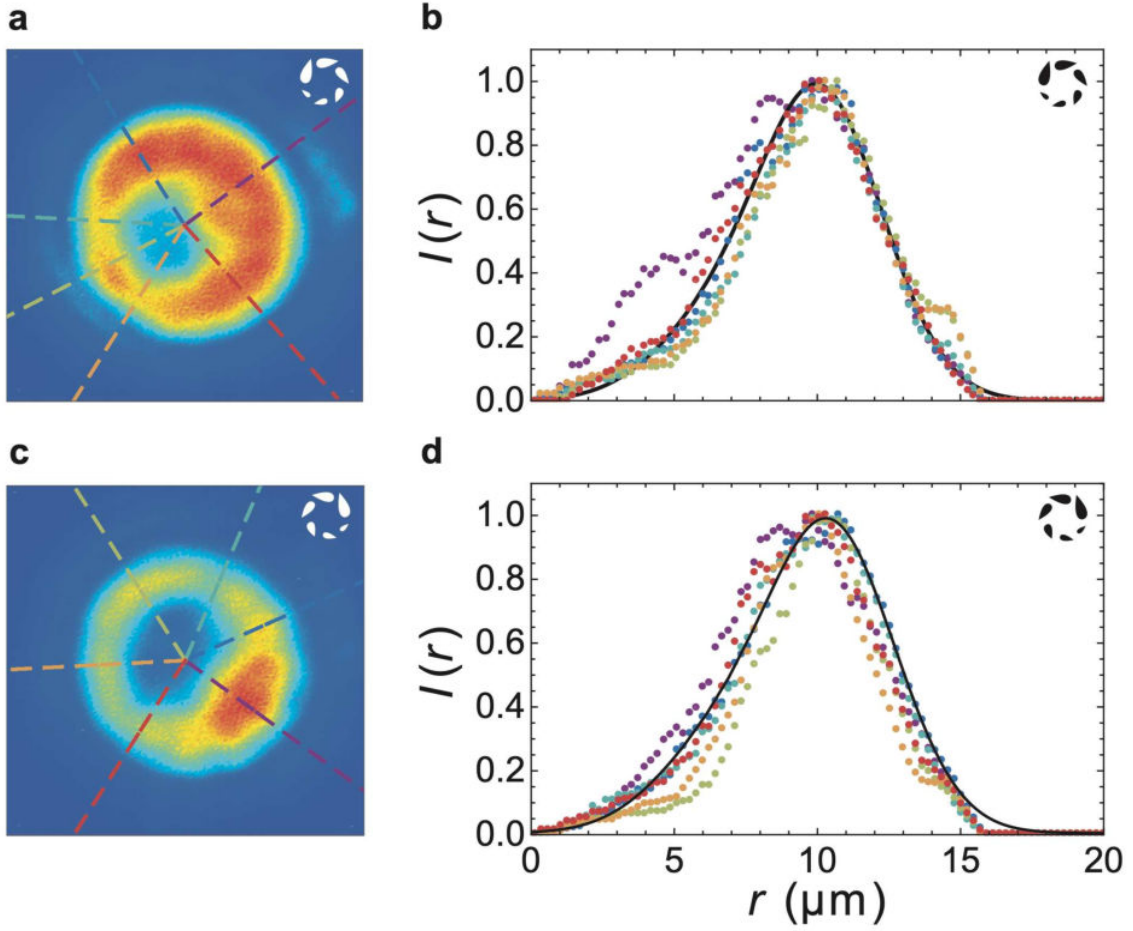

Figure S1: The radial distribution of the normalized intensity of the half-moon polariton condensates. (a), (c) The intensity distribution in the cavity plane and (b), (d) the radial intensity distribution of the polariton current states with the positive [(a), (b)] and negative [(c), (d)] OAM. Dashed color lines in (a) and (c) indicate the cross-sections in which the intensity distribution is shown in (b) and (d) in the same color, respectively. The black curves in (b) and (d) are the best fits of the average radial distribution of the condensate by the analytical function $(e + dr + cr^2)e^{-a(r-b)^2}$. The coefficients in the fitting function are as follows: $a = 0.0784\,\mu m^{-2}$, $b = 8\,\mu m$, $c = 0.0467\,\mu m^{-2}$, $d = -0.5334\,\mu m^{-1}$ and $e = 1.9778$ for (b) and $a = 0.0794\,\mu m^{-2}$, $b = 8.090\,\mu m$, $c = 0.0393\,\mu m^{-2}$, $d = -0.4148\,\mu m^{-1}$ and $e = 1.5144$ for (d). The roundel of droplets in the upper left corner of panels indicates the polariton current state: counter-clockwise for $\ell > 0$ and clockwise for $\ell < 0$.

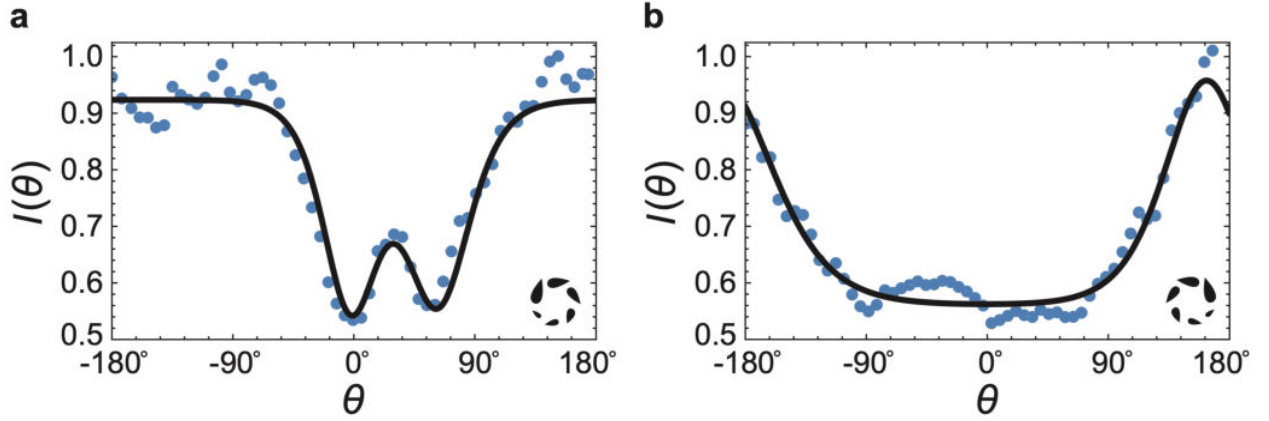

Figure S2: The azimuthal distribution of the normalized intensity of the half-moon polariton condensates. (a) The intensity distribution of the positive and (b), the negative OAM polariton current state. Blue dots denote the experimental measurements, black curves show the fitting by the analytical function $a\tanh{[b(\theta+c)]}^2 + a\tanh{[d(\theta-e)]}^2 + f$. The fitting parameters are estimated as follows: $a = 0.3574$, $b = 2.1093$, $c = 0.1289$, $d = 1.8003$, $e = 1.0064$, $f = 0.2079$ for (a) and $a = -0.5433$, $b = 1.0885$, $c = 3.4725$, $d = 1.3705$, $e = 2.8511$, $f = 1.4708$ for (b). The angle $\theta = 0°$ corresponds to the position of the phase jump (see Fig. S3).

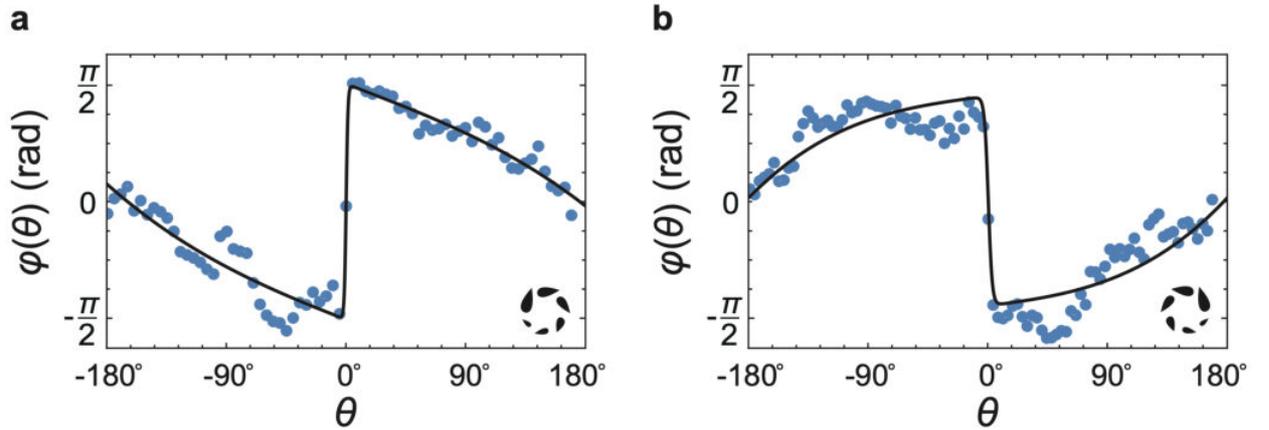

Figure S3: The azimuthal distribution of the phase of the half-moon polariton condensates. (a) The phase distribution of the positive and (b), the negative OAM polariton current state. Blue dots denote the experimental measurements, black curves show the fitting by the analytical function $a\theta + b\theta^2 + c\theta^3 + d\tanh(e\theta)$. The fitting parameters are estimated as follows: $a = 0.1361$ $b = 0.0047$, $c = 0.0321$, $d = -1.4335$, $e = 19.7694$ for (a) and $a = -0.4032$, $b = 0.0092$, $c = -0.0155$, $d = 1.606$, $e = 41.7494$ for (b).

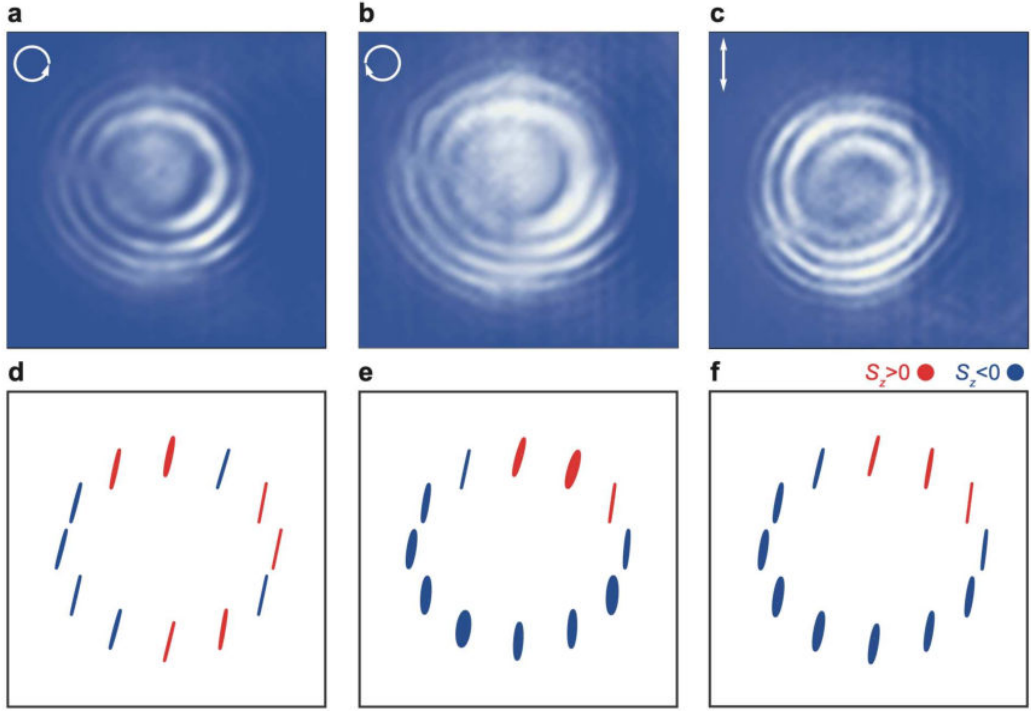

Figure S4: The interferometry images (upper panels) and the polarization distributions in the near field emission spectra of the half-moon condensates (lower panels) shown by the polarization ellipses placed at the various positions along the condensate ring for the condensates excited by a non-resonant optical pump of the (a), (d) left-circular, (b), (e) right-circular and (c), (f) linear polarizations. White arrows in the upper left corners of the upper panels indicate the polarization of the pump beam. Red and blue ellipses in the lower panels correspond to $S_z > 0$ and $S_z < 0$, respectively. The data are shown for the counter-clockwise polariton current states.

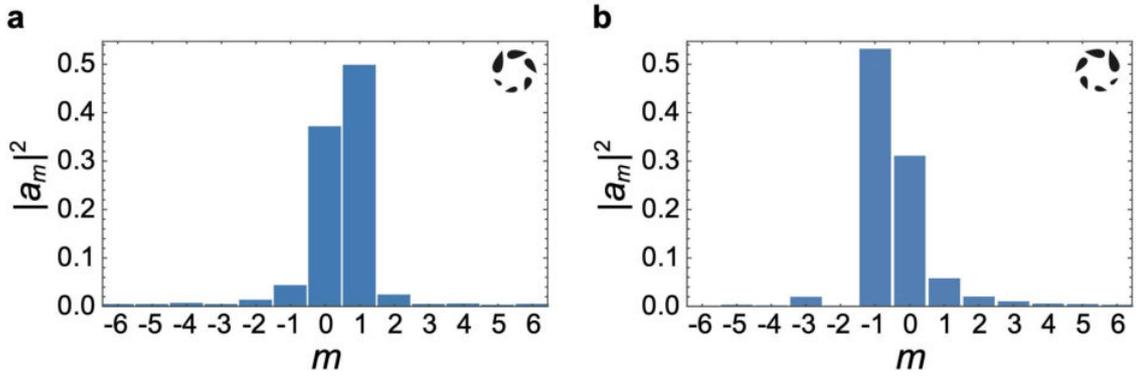

Figure S5: Contribution $|a_m|^2$ of the current states with the winding numbers $m$ to the half-moon condensates. Panels (a) and (b) are for the counter-clockwise and clockwise polariton current states, respectively.

6